\documentclass[12pt]{iopart}
\usepackage{graphicx}

\begin{document}

\title[Experimental study on controlled production of two-electron temperature plasma]{Experimental study on controlled production of two-electron temperature plasma}

\author{G. Sharma$^{1}$, K. Deka$^{1}$, R. Paul$^{1}$, S. Adhikari$^{2}$ R. Moulick$^{3}$, S. S. Kausik$^{1,*}$, and B. K. Saikia$^{1}$}

\address{$^1$Centre of Plasma Physics, Institute for Plasma Research, Nazirakhat, Sonapur-782402, Kamrup(M), Assam, India}
\address{$^2$Department of Physics, University of Oslo, PO Box 1048 Blindern, NO-0316 Oslo, Norway}
\address{$^3$Department of Physics, Rangapara College, Rangapara-784505, Sonitpur, Assam, India}
\address{$^*$E-mail: kausikss@rediffmail.com}
\vspace{10pt}

\begin{abstract}
A two-electron temperature plasma is produced by the method of diffusion of two different plasmas with distinct temperatures and densities. The method is simple and provides an adequate control over the plasma parameters. The study reveals that the temperature and density of both the electron groups can be effectively controlled by just changing the discharge currents of both the plasmas. An ion-acoustic (IA) wave is excited in the plasma and is detected using a planar Langmuir probe. The damped amplitude of the wave is measured and is used as a diagnostic tool for establishing the presence of two-electron components. This production method can be helpful in controlling the hot electron density and temperature in plasma processing industries.
\end{abstract}

%
%
%
%
%

\section{Introduction}
The study of two-electron temperature plasma is a stimulating field of research due to its substantial properties and underlying intricacies. It is frequently found in laboratories as well as in space environment. For example, thermonuclear plasmas often have a high energy tail\cite{Quon,Stan}, plasmas in a double plasma device \cite{Arms} and usual hot cathode discharge plasma\cite{Naka} can also have two different electron groups. On the other hand, plasma in the earth's magnetosphere\cite{Coro} and Saturn's ring\cite{Young} are examples of naturally occurring plasmas having two-electron components. In the case of laboratory plasmas, both the electron components are often characterized by Maxwell distribution. But, due to the presence of an extremely high energy electron component, also known as superthermal electrons in space plasmas, they are far from being Maxwellian. Instead of an exponential decrease, they often follow a power-law decrease which is precisely characterized by the kappa-distribution\cite{Schi}. The characteristics of such a plasma are found to be different than that of a normal electron-ion plasma. For example, the presence of energetic electrons is responsible for propagation of electron acoustic waves\cite{Berth, Gary,Ike} in plasmas. Also, it supports both positive and negative ion-acoustic soliton propagation, whereas negative potential disturbances can not sustain in a normal electron-ion plasma\cite{Baluku}. The production of negative ion is one of the major research problem related to fusion experiments. Tobari \textit{et. al.}\cite{Tobari} experimentally observed that high energy electrons have a significant effect on negative ion production in hydrogen plasma when a small amount of cesium is seeded into the system. On the other hand, in magnetron plasmas, the presence of a hot electron group introduces a noteworthy effect on sputtering yield\cite{Haase}. Hence, it becomes important to carry out more experimental studies to understand the behaviour of a second electron group in plasma.

The review of the literature reveals that two groups of electrons having different energies and densities are prevalent in experimental devices. In low pressure plasmas ($10^{-5}$ $mbar$), three groups of electrons are observed\cite{Rus, Mishra}, \textit{viz.}, ionising electrons, hot electrons and cold electrons. The ionising electrons are emitted from the filaments, which are accelerated by the discharge voltage towards the chamber walls. Such electrons suffer collisions with the background gas and lose energy and form the hot electron group. On the other hand, cold electron group is formed due to the ionisation of neutral gas species. At high neutral pressure, the ionising electrons become identical to that of the hot electrons due to an increase in inelastic collisions with the neutrals and consequently results in a bi-Maxwellian electron distribution\cite{Mishra,Chan}. It has been found that despite their frequent occurrence in multi dipole devices, there are very few general techniques available for controlling their density as well as temperature. It turns out to be a limitation for such experimental devices, where both electron components play a vital role. Yamazumi and Ikezawa\cite{Yama} had reported such a parameter control mechanism for two-electron temperature in a double plasma (DP) device. But their method is only useful when plasma potential is close to filament potential. Pustylink \textit{et al.}\cite{Pusty} had also carried out studies in controlling energetic electrons in DP device. Their technique is based on controlling the anode potential and the study was focused on the plasma obtained in the target chamber. Some other experimental studies\cite{Sato,Kato,Kato2} had also been carried out in this regard but these were primarily focused on fulfilling some specific purpose, which can not served as a general treatment. In these studies, electron temperature is mainly controlled by introducing various grids in the plasma chamber. On the other hand, Hass \textit{et al.} had reported that in a capacitive discharge, presence of a mono energetic electron beam can tailor the hot tail of electron distribution function\cite{Haa}.  Mishra and Phukan\cite{Mishra} had studied two-electron temperature plasma using single magnetic cage and reported a control mechanism for electron density and temperature by biasing the magnetic cage used for plasma confinement. The method is helpful in changing the hot electron temperature significantly but the changes observed in their density is really small. Hence, a method for controlling the electron energies as well as densities in a two electron temperature plasma is necessary which can be generalized for any experimental devices. Therefore. the primary focus of the present work is to develop a simple and efficient technique for production and control of two-electron temperature plasma. The work is a continuation of the earlier theoretical study of sheath properties in a magnetized two temperature plasma\cite{My} of this group. To support the theoretical results, the development of a simple and controlled production technique of two temperature plasma is the main motivation behind the present study. In this paper, a simple method of production of two-electron temperature plasma is described. It can boost experimental studies in two temperature plasma because of its simplicity. To verify the presence of two-electron groups, Langmuir probe and ion-acoustic wave are used as diagnostic tools. The paper is organized as follows. In section \ref{Expt}, the details of the experimental arrangement are described. Section \ref{R&D} contains the experimental results with discussions. And in section \ref{Conc}, a concluding remark is given.

\section{Experimental set up}\label{Expt}
\begin{figure}
    \centering
    \includegraphics[width=0.5\textwidth]{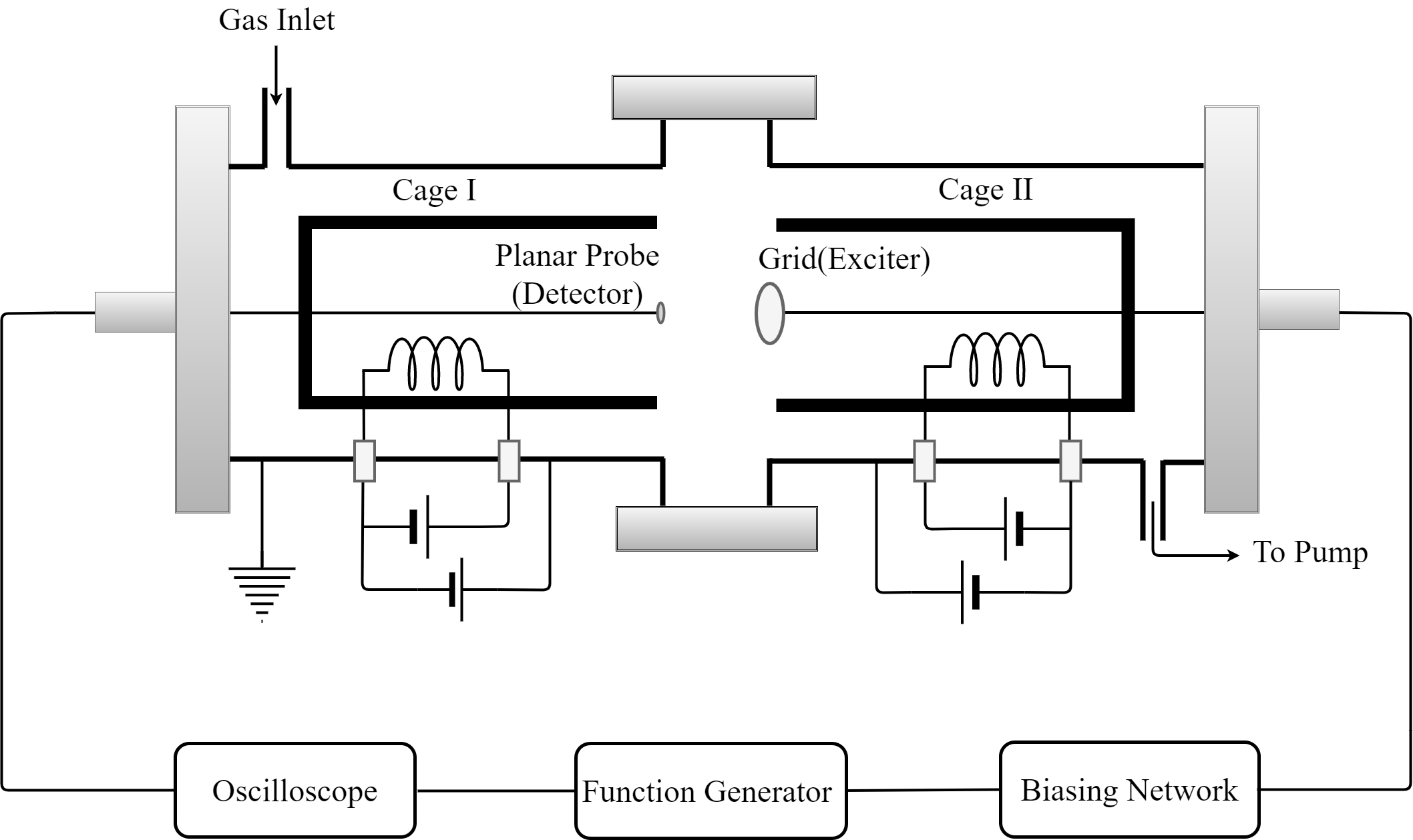}
    \caption{Schematic of the experimental set up}
    \label{fig:1}
\end{figure}
The experiment is performed in a cylindrical stainless steel chamber of 100 cm length and 30 cm in diameter. The schematic of the experimental arrangement is shown in figure \ref{fig:1}. Usually, plasma is produced in this chamber using hot cathode discharge mechanism, where the discharge strikes due to ionisation of the background gas by the primary electrons emitted from the filaments and a full line cusp magnetic cage is used for plasma confinement. The plasma produced in this manner is found to be Maxwellian in nature with a single electron component. To obtain two-electron components, two cusp magnetic cages are used. These magnetic cages have different surface field strengths and different confining properties. For convenience, they are labelled as cage I and cage II.

Cage I is made up of samarium cobalt permanent magnets of surface field strength 3.5 kG. Water cooling for the cage is necessary during operation. Due to its higher field strength compared to the other cage, high-density plasma is obtained ($10^{17}-10^{18}$ $m^{-3}$). It can sustain up to a discharge current of 10 A. On the other hand, cage II is made up of strontium ferrite permanent magnets of surface field strength 1.2 kG. Due to its lower field strength, a comparatively low-density plasma is obtained ( nearly $10^{16}~m^{-3}$). The cage is usually operated with a discharge current less than 1 A. 

Four power supplies are used in the whole plasma production process. Two power supplies are (filament and discharge) dedicated for
each cage. The discharge power supplies are always operated at 80 V. In cage I, the filament voltage is varied between 10 V to 14 V with a filament maximum current 22 A. On the other hand, in cage II, the filament voltage is varied between 14 V to 20 V with a filament current of maximum 6 A. In such conditions, the discharge current, in cage I, reaches a maximum of 4 A and in cage II, reaches a maximum of 1 A.

\subsection{Production of two electron temperature plasma}
The main idea behind the technique is to diffuse two plasmas of different characteristics. As already mentioned, two plasmas are produced separately and are allowed to diffuse in the centre of the chamber. A constant pressure is maintained throughout the chamber and a common gas species is used. After production, plasma from both the cages quickly diffuses with each other. The region where both the plasmas are allowed to diffuse is a magnetic field-free region; therefore, the magnetic field does not affect particle diffusion in this case. Now to get an idea of the nature of diffusion taking place, the electron free diffusion force $F_D$ and diffusion force due to space charge electric field $F_E$ has been estimated using the following approximate relations\cite{Bitt}.
\begin{equation}\label{fd}
    F_D = \frac{nT_e}{m_en_eL}
\end{equation}
\begin{equation}
    F_E = \frac{e^2nL}{m_e\epsilon_0}
\end{equation}
where $n$ is the electron density difference, $T_e$ is the effective electron temperature, $L$ is the characteristic length over which density varies significantly, $m_e$ is the electron mass, $e$ is charge of a electron and $\epsilon_0$ is the permitivity of free space. The effective electron temperature can be given as\cite{Song,Mer} 
\begin{equation}
    T_e = (N_c+N_h)/((N_c/T_c)+(N_h/T_h))
\end{equation}
where $N_c$ and $N_h$ are cold and hot electron densities and $T_c$ and $T_h$ are corresponding electron temperatures, respectively. For the present experimental conditions, it has been found that $F_E>>F_D$. Hence, it is understood that the ambipolar electric field dominates due to which diffusion takes place in the system. 

It has been found that the full line cusp magnetic field configuration helps in primary electron confinement\cite{Taylor,Limpa,Bosch}. As a result, often the electron energy distribution is observed with a hot tail. Here, the cage I provides better confinement over cage II. Hence, a significant number of hot electrons will be supplied by cage I. During single cage operation, mean electron temperature of the cage I is found to be less than 10 eV and that of cage II is found to be around 2 eV depending on working pressure and applied voltage. To ensure that the difference in energy as well as in density between both the plasmas is maintained, the filaments used in the cages have different surface areas. A tungsten filament of length 15 cm and diameter 0.5mm is used in the cage I and another filament of the same length but of diameter 0.2mm is used in cage II.

It is found that for certain combinations of the discharge currents, two distinct electron groups are observed with two different temperatures. The distinction in the temperatures is more prominent when cage I has a higher discharge current in comparison with cage II.

In the experiment, the gas is injected from one end of the chamber, and the pumping is done from the opposite end. The experiment was first performed using cage I towards the outlet and cage II towards the gas inlet. It has been seen that interchanging the position of the cages does not affect the plasma characteristics.

\subsection{Excitation of ion-acoustic wave}
After completing the first part of the experiment, a stainless steel mesh grid is inserted in the chamber along with a planar probe to launch ion-acoustic (IA) waves in the system. The circular grid has a diameter of 5 cm with 60\% grid transparency. On the other hand, the planar probe has a diameter of 9 mm.
The grid and the probe are placed in the central region of the plasma with a maximum separation of 10 cm where hot and cold both electron components are present. Now, using a function generator a sinusoidal voltage pulse is applied to the grid using a biasing network and the applied pulse is detected by the planar probe. It is seen that the detected wave has considerably low amplitude which indicates the damping of the applied pulse. Now the reason behind the damping may be either of collisional origin or due to wave-particle interaction known as Landau damping. But as the electron temperature of the plasma is much higher than that of ion temperature, therefore collisional damping will be the dominating factor in this case.

\subsection{Plasma diagnostics}
\begin{figure}
    \centering
    \includegraphics[width=0.5\textwidth]{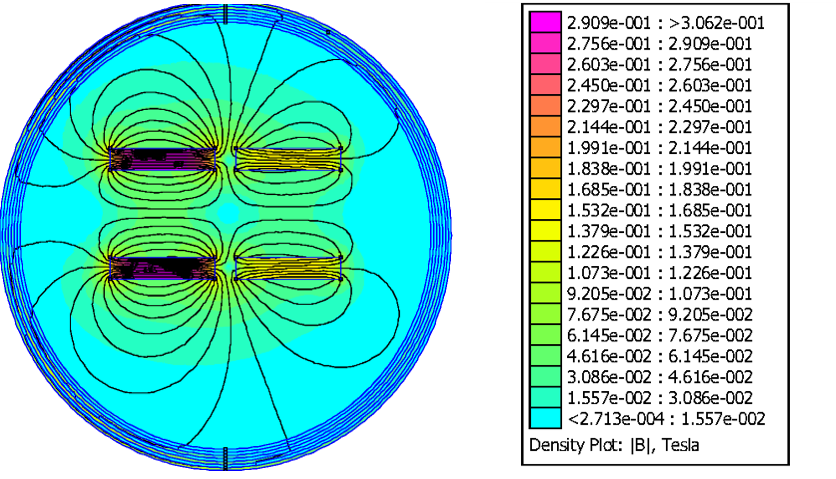}
    \caption{Resultant magnetic lines of force inside the chamber in $r-z$ plane}
    \label{fig:2}
\end{figure}

The general theory of Langmuir probe is applicable for an unmagnetized plasma\cite{Kalita}. In a full line cusp magnetic cage, the axial region is field-free, and hence the Langmuir probe measurement is valid. In the present case, one end of both the cages in the centre of the chamber is kept open for plasma diffusion. Hence, it is of utmost importance to know the orientation of magnetic field lines in that particular volume of the chamber. Figure \ref{fig:2} shows the orientation of magnetic lines of force in a vertical plane. It is confirmed that the central region of the chamber is also field free. An EPSION advanced cylindrical Langmuir probe (Hiden Analytical) is used for determination of plasma parameters. The tungsten probe tip has a diameter of 0.3 mm and a length of 3 mm. The chamber is pumped down to a base pressure of $5\times10^{-6}~mbar$ using the combination of a rotary and diffusion pump. A constant working pressure of $2\times10^{-4}~mbar$ is maintained throughout the experiment.

\section{Results and Discussions}\label{R&D}
\subsection{Observation of two electron groups}
\begin{figure}
    \centering
    \includegraphics[width=0.45\textwidth]{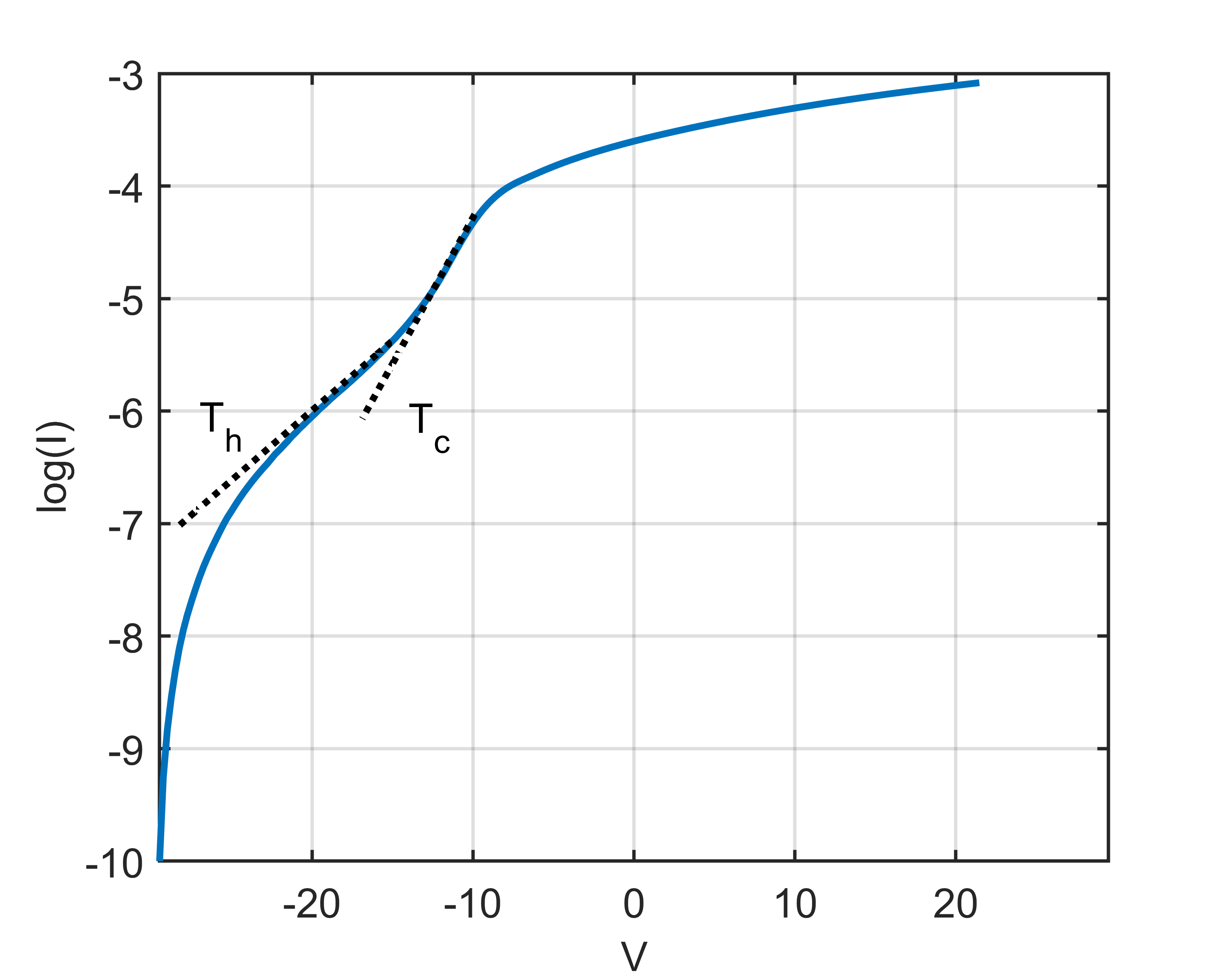}
    \caption{Semi-logarithmic \textit{I-V} characteristic showing the presence of two electron temperatures}
    \label{fig: 3}
\end{figure}

\begin{figure}
    \centering
    \begin{minipage}[b]{0.45\textwidth}
    \includegraphics[width=1\textwidth]{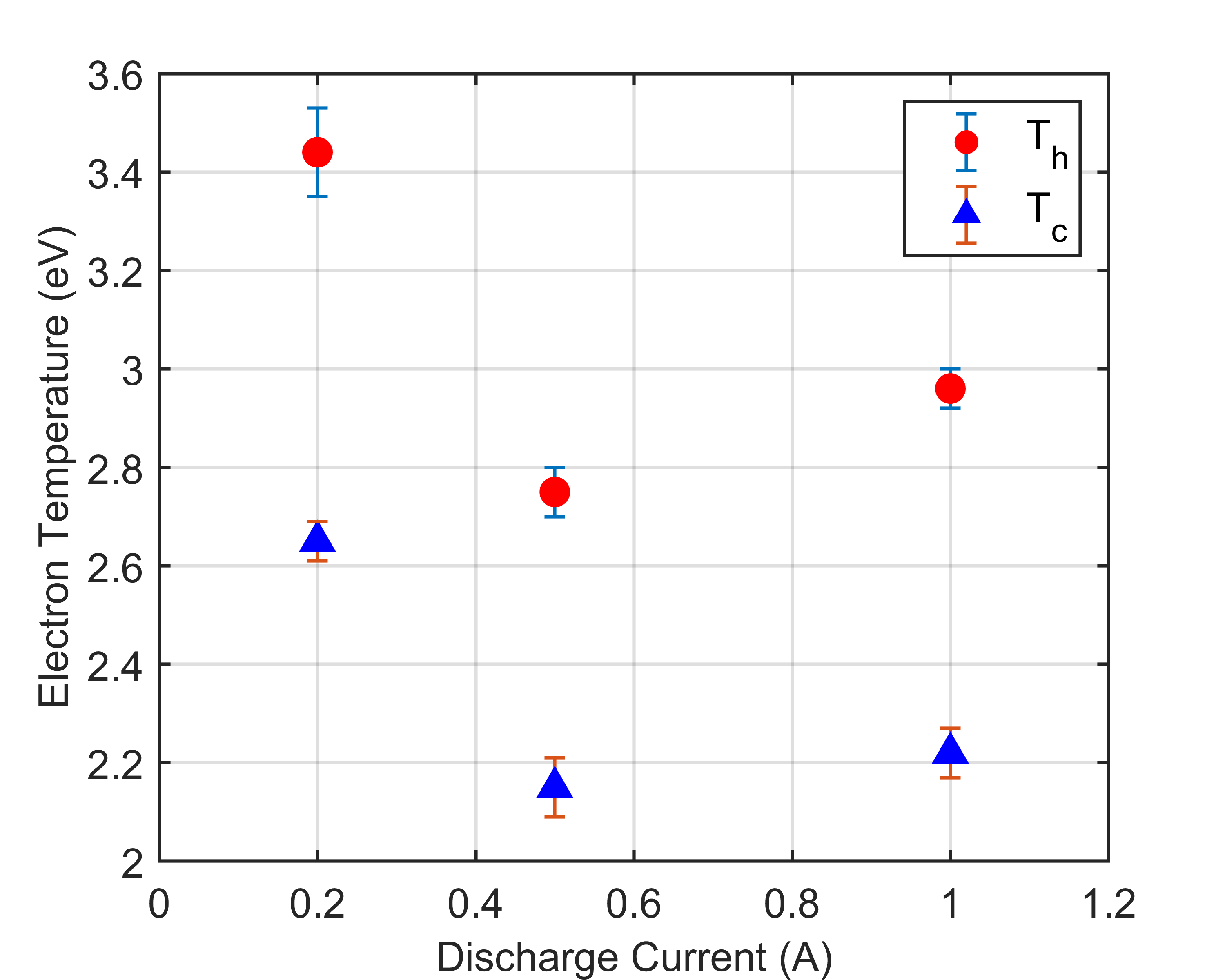}
    \caption{Electron temperature measurement for constant discharge current of 2 A in cage I and variable discharge current in cage II}
    \label{fig: 4}
    \end{minipage}
    \begin{minipage}[b]{0.45\textwidth}
    \includegraphics[width=1\textwidth]{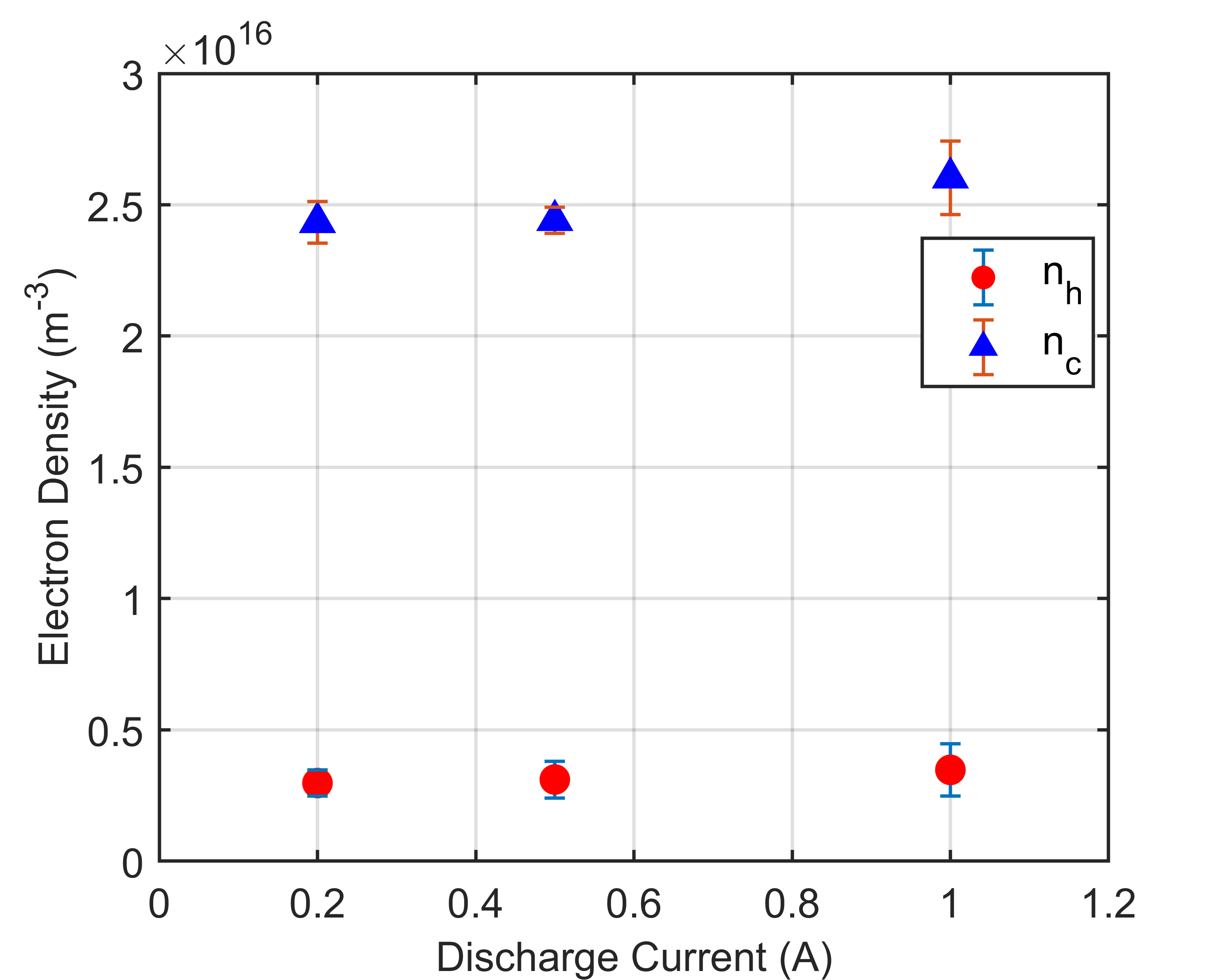}
    \caption{Electron density variation for constant discharge current of 2 A in cage I and variable discharge current in cage II}
    \label{fig: 7}
    \end{minipage}
\end{figure}

To observe a two-electron population, the discharge currents in both the cages are varied. In other words, the applied voltage to the filaments is changed. The Langmuir probe gives the \textit{I-V} characteristics of the plasma and the corresponding electron temperature, electron density and electron energy distribution function (EEDF) is calculated. At first, the discharge current in the cage I is maintained at 2 A and varied in cage II from 0.2 A to 1 A. In this case, since discharge current in cage II cannot be increased beyond 1 A; hence three values of discharge current has been chosen where an appreciable change in electron temperature is observed. The semi-logarithmic plot of voltage vs. current characteristic has two distinct slopes indicating the presence of two-electron groups (figure \ref{fig: 3})\cite{Pusty,Pil}. The experiment has been performed using three different gas species, \textit{viz.,} hydrogen, nitrogen and argon. The results only for hydrogen plasma has been portrayed in the manuscript as the results for the other two gases also show similar trend. The values of electron temperatures and electron densities are calculated for three different discharge currents in cage II. Figure \ref{fig: 4} represents the electron temperature variations and the corresponding density variations of electron groups are also calculated as shown in figure \ref{fig: 7}. It is observed that electron temperature decreases with increase in discharge current. The decrease in electron temperature is also reflected  in the floating potential of corresponding \textit{I-V} characteristics. An increase in discharge current in cage II consequently increases the plasma density whereas the plasma density in cage I is kept constant my maintaining a fixed discharge current of 2 A. This results in decrease of density and temperature gradient between the cages and as a consequence of thermalization, the electron temperature gradually decreases.
\begin{figure}
    \centering
    \begin{minipage}[b]{0.45\textwidth}
    \includegraphics[width=1\textwidth]{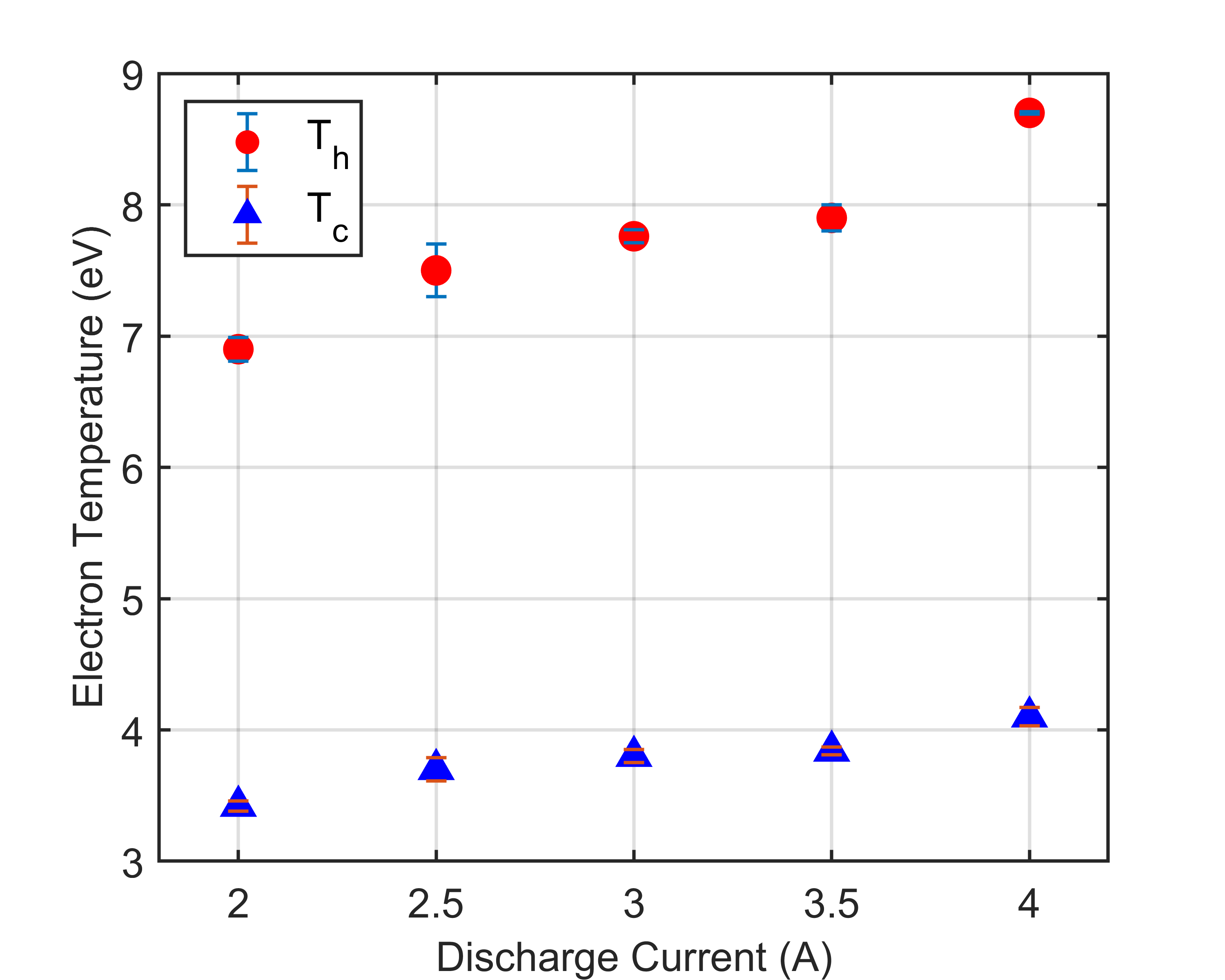}
    \caption{Electron temperature measurement in combined cage configuration with constant discharge current of 0.2 A in cage II and variable discharge current in cage I}
    \label{fig: 8}
    \end{minipage}
    \begin{minipage}[b]{0.45\textwidth}
    \includegraphics[width=1\textwidth]{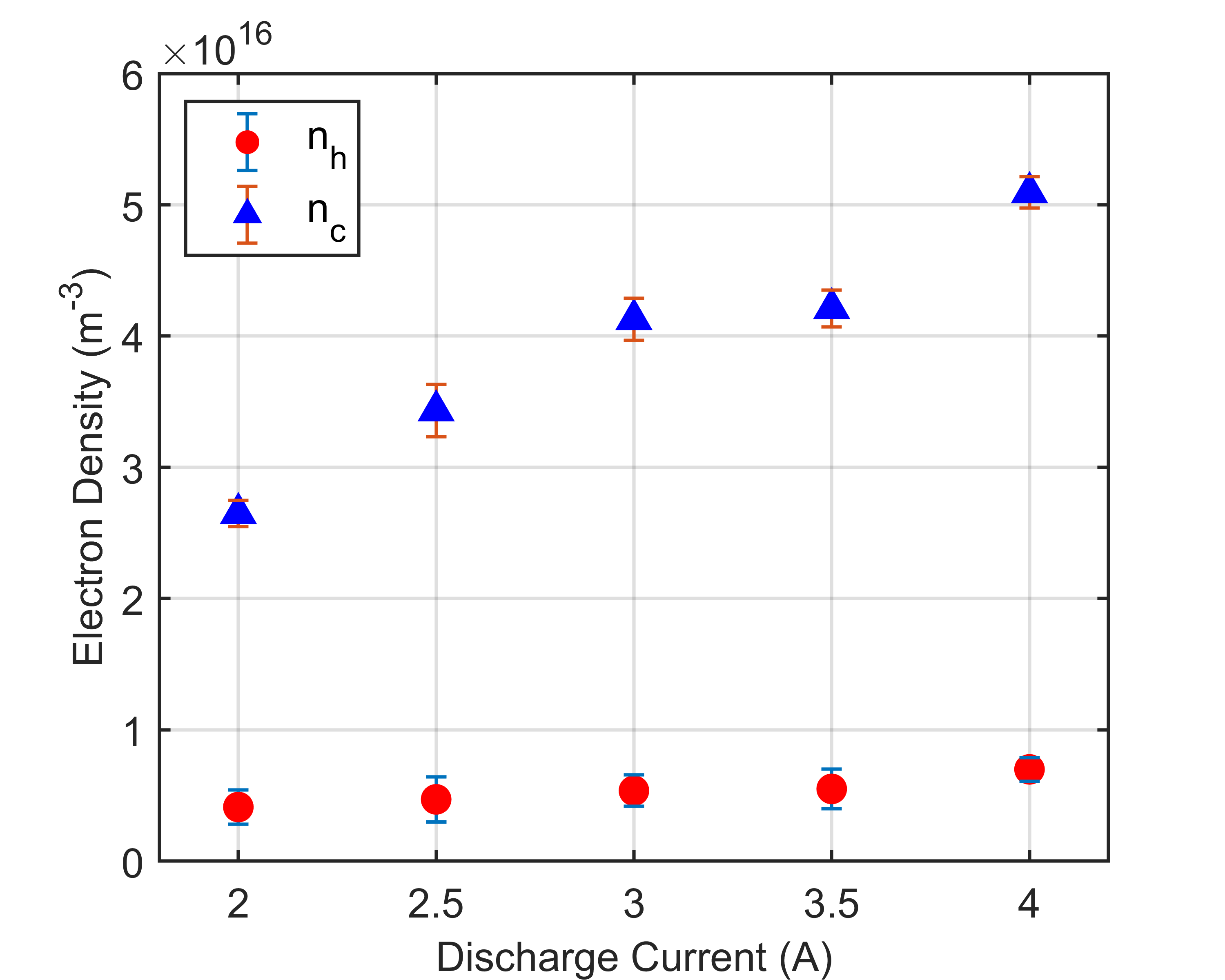}
    \caption{Electron density variation in combined cage configuration with constant discharge current of 0.2 A in cage I and variable discharge current in cage I}
    \label{fig: 11}
    \end{minipage}
\end{figure}

Now, the discharge current in cage II is kept constant at 0.2 A and current in the cage I is varied from 2 A to 4 A in the step of 0.5 A. The variations observed in the electron temperature is shown in figure \ref{fig: 8}. In this case, the temperature of both the electron groups is observed to increase with the increase in the discharge current. This is because, increase in the discharge current introduces more energetic primary electrons in the system, which in turn generate ion-electron pairs by ionizing the background gas molecules. The corresponding densities for both the electron groups have been calculated and found an increase in the cold electron density whereas hot electron density approximately remains same with discharge current. The hot and cold electron densities for hydrogen discharge are shown in figure \ref{fig: 11}.

\begin{figure}
    \centering
    \includegraphics[width=0.45\textwidth]{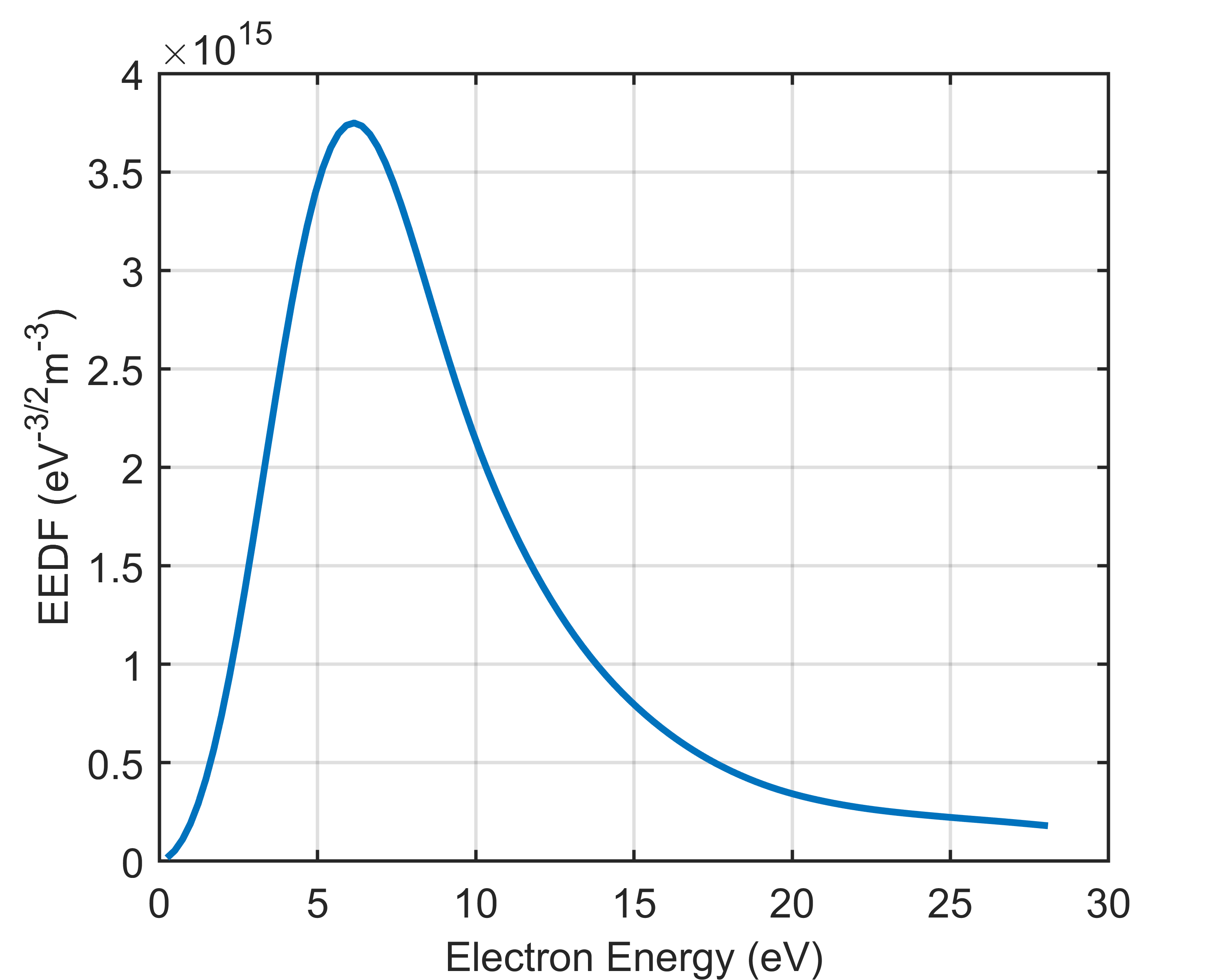}
    \caption{EEDF measurements for a regular discharge in a single cage operation}
    \label{fig: 14}
\end{figure}

\begin{figure}
    \centering
    \begin{minipage}[b]{0.45\textwidth}
    \includegraphics[width=1\textwidth]{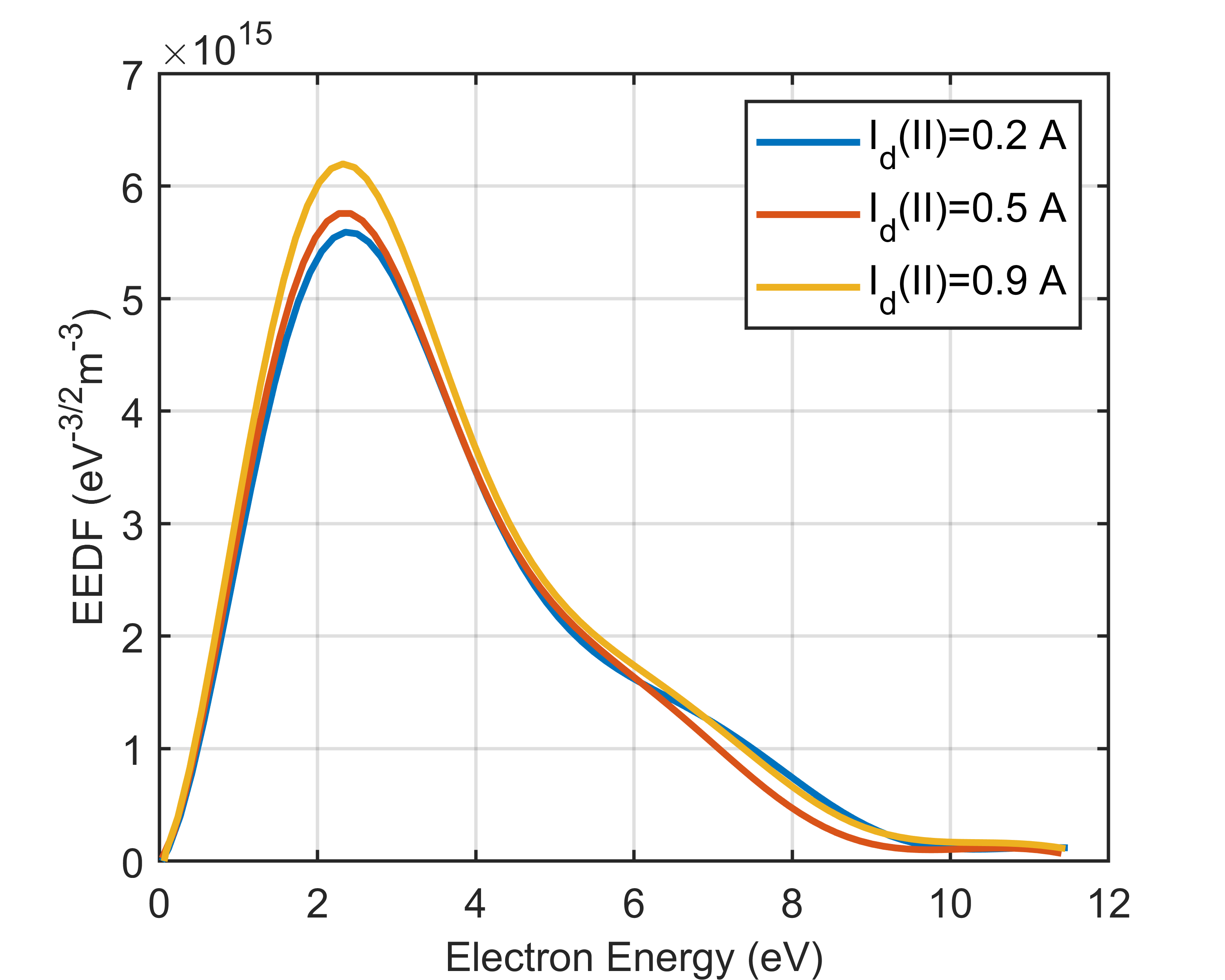}
    \caption{EEDF measurements for combination of cages with constant discharge current of 2 A in cage I}
    \label{fig: 12}
    \end{minipage}
    \begin{minipage}[b]{0.45\textwidth}
    \includegraphics[width=1\textwidth]{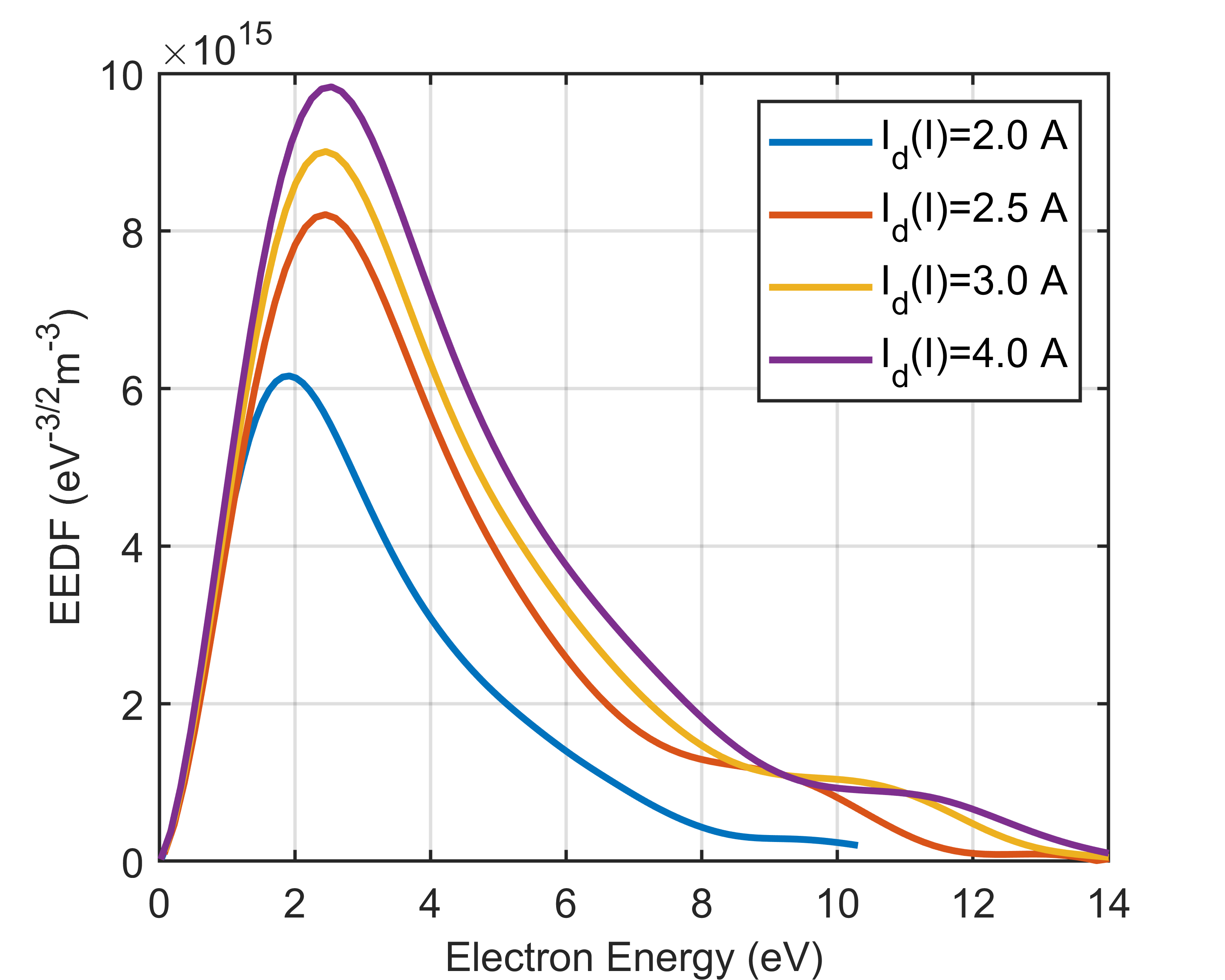}
    \caption{EEDF measurements for combination of cages with constant discharge current of 0.2 A in cage II}
    \label{fig: 13}
    \end{minipage}
\end{figure}

Figure \ref{fig: 12} shows a typical EEDF plot for a plasma produced in a single cage. One can notice that figure \ref{fig: 12} is of Maxwellian nature. Figure \ref{fig: 13} represents the EEDFs for three different discharge currents in cage II with a fixed discharge current in the cage I. Here, it is visible that all the three curves have a primary peak at around 2.5 eV and a second hump at about 5 eV. This second hump corresponds to the other group of electrons having comparatively higher temperature but less density. On the other hand,  Figure \ref{fig: 14} shows the corresponding EEFDs for constant discharge current in cage II and variable discharge current in the cage I. Since cage I can produce a high-density plasma with a high electron temperature, increasing the discharge current in the cage I introduces more energetic electrons in the system. These electrons then diffuse with the plasma electrons from cage II and thermalize themselves at a higher temperature. The EEDFs in figure \ref{fig: 14} has an energetic tail with a small bump at around 10 eV. This corresponds to the ionising electrons whose energy is lost due to collisions with the neutrals. The little bump on the tail becomes prominent and gradually shifting to higher energy side for higher discharge currents indicating the increase of primary electrons emitted from the filaments. Also, a comparison between figure \ref{fig: 13} and figure \ref{fig: 14} reveals that to increase the energy separation between the electron groups, the discharge current in the cage I has to be increased keeping a small discharge current in cage II.

\begin{figure}
    \centering
    \includegraphics[width=0.45\textwidth]{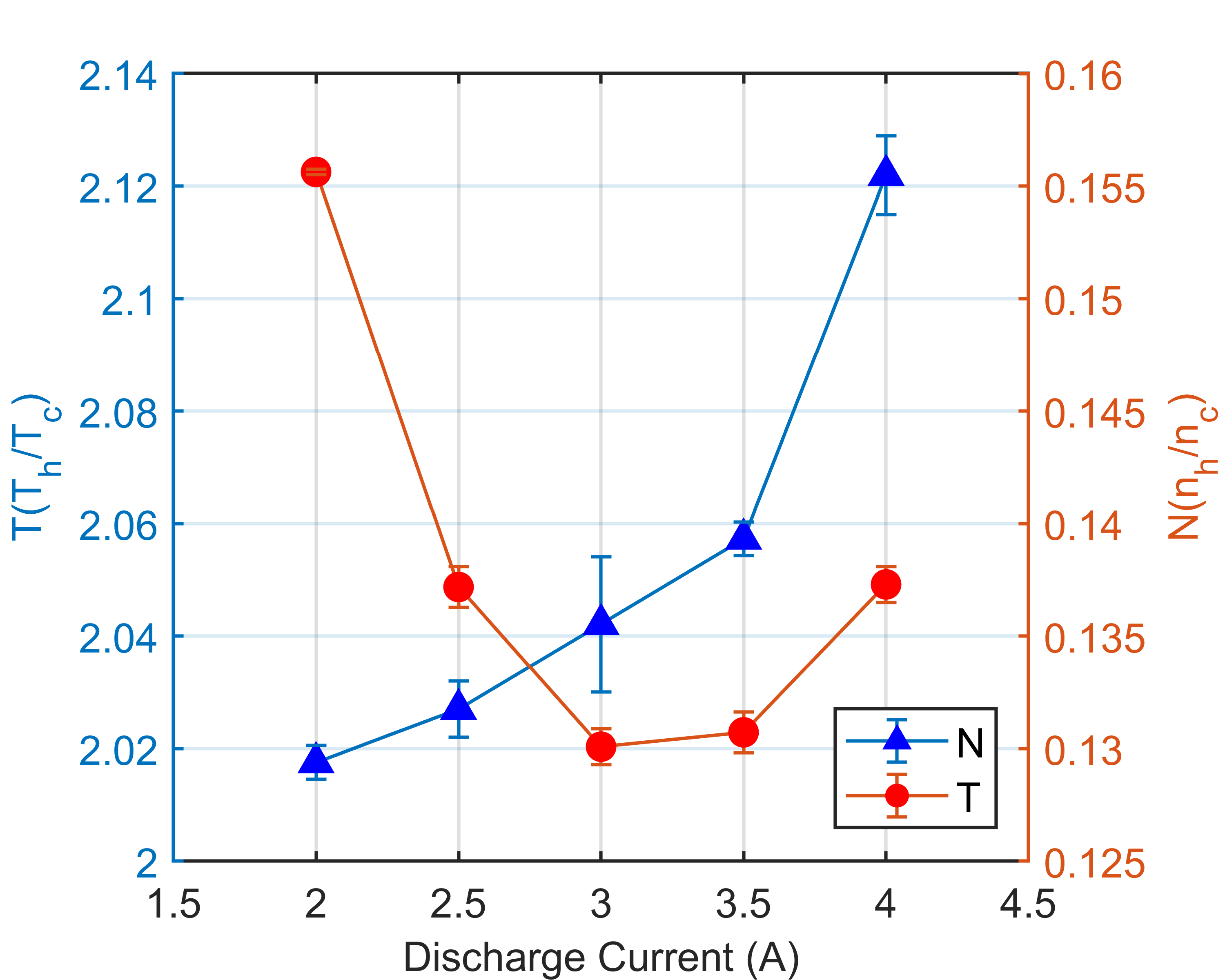}
    \caption{Ratio of hot to cold electron density and temperature}
    \label{fig:15}
\end{figure}

The ratio of hot to cold electron density ($N=n_h/n_c$) and temperature ($T=T_h/T_c$) is also found out and plotted in Figure \ref{fig:15}. It has been observed that the electron temperature ratio increases with the increase in discharge current but density ratio decreases first, attains a minimum and then starts rising again. Looking at the electron density variations in figure \ref{fig: 11}, it can be understood that the decrease in $N$ is attributed to increasing in $n_c$ in comparison to that of $n_h$. From these density and temperature ratio plots, it is observed that increasing the discharge current in cage I and maintaining a low discharge current in cage II, the temperature of the hot electrons can be increased further. In contrast, their fractional density can be significantly decreased.

Thus, by playing with the discharge currents in both the cages, the desired electron energy and density can be achieved. To increase the hot electron density, the cage I can be replaced with another cage of higher surface field strength. Moreover, to increase the hot electron temperature, more number of filaments can be used. But one significant observation may be worth mentioning at this stage. It is observed that, if primary electron density in the cage I is made very high by using more number of filaments compared to cage II, the plasma becomes very unstable. It may be due to the occurrence of sharp density gradient across the junction.

The theoretical study on sheath formation in a collisional two-electron temperature plasma\cite{My} suggests that the presence of hot electrons creates detrimental effects in plasma processing. Since the present technique provides a controllable mechanism for producing two electron groups, if the plasma is used for surface modification where maximum ion flux is required, this method becomes handy in reducing the hot electron population. On the other hand, wherever dual electron traits are desired, the same method can be easily adopted.

\subsection{Diagnosis by ion-acoustic wave}

\begin{figure}
    \centering
    \includegraphics[width=0.45\textwidth]{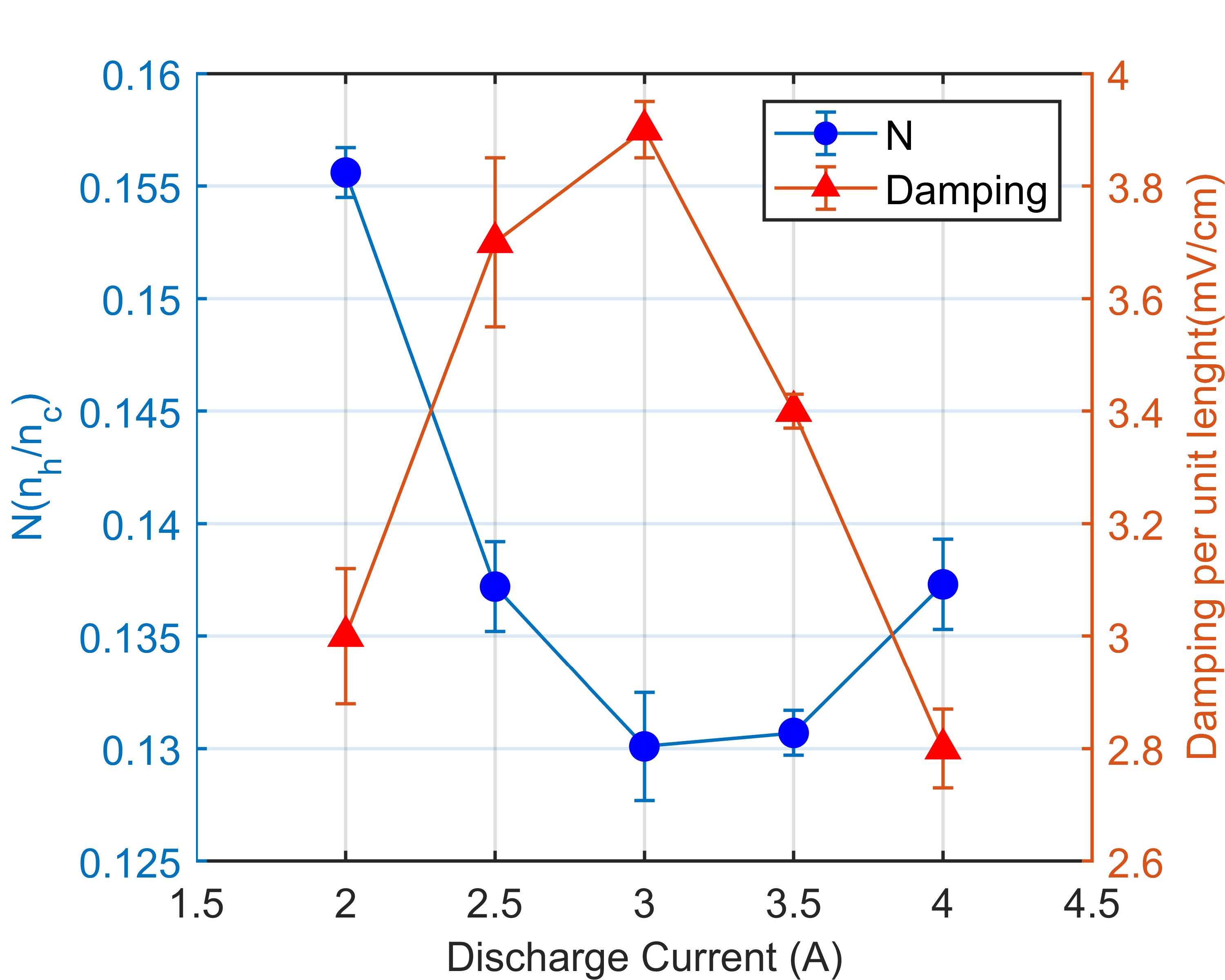}
    \caption{Experimentally measured damping per unit length of IA wave and corresponding hot to cold electron density ratio for different discharge currents}
    \label{fig:18}
\end{figure}

To further verify the results obtained from Langmuir probe measurement, ion-acoustic wave damping is used as a diagnostic tool for confirmation of the presence of two-electron components. Theoretical studies have shown that the presence of a hot electron group modifies the ion-acoustic waves propagation characteristics in plasma\cite{Baluku,Moslem,Ver}. In a kinetic simulation of IA wave propagation in two-electron temperature plasma, it has been found that the presence of a significant population of hot electron enhances the growth of IA wave. In contrast, the decrease in the effective electron temperature contributes to damping\cite{Koen}. To observe the effect of hot electrons, an ion-acoustic wave is excited in the system, the details of which have already been described in section \ref{Expt}. The applied wave has a peak to peak voltage of 3 V, and the frequency is 50 kHz. This particular experiment is carried out in hydrogen plasma by keeping 0.2 A discharge current in cage II and varying the current in the cage I. The distance between the exciter and the receiver is varied from 2 cm to 10 cm. The applied signal to the exciter is detected by the receiver and found to be damped. For five different values of discharge currents, the relative damping of the signal is calculated. These five discharge currents correspond to five different sets of hot and cold electron densities. When the damping per unit length is plotted against different discharge currents as shown in figure \ref{fig:18}, it is found that maximum damping is observed for discharge current of 3 A which corresponds to the lowest value of hot to cold electron density ratio $N$. This particular observation supports what is shown in figure \ref{fig:15}. It has been found that with the increase in discharge current, the ratio $N=n_h/n_c$ decreases and reaches a minimum at 3 A, indicating an increase in cold electron density. As shielding mainly depends on the density of the cold electron component, hence decreasing $N$ consequently favours the damping.

\section{Conclusion}\label{Conc}

In summary, a simple but effective technique is developed for the production of two-electron temperature plasma in the laboratory. The method can be implemented to study different phenomena in plasma, \textit{e.g.}, sheath formation, wave propagation, dust charging in the presence of two-electron groups. It gives reasonable control over the plasma parameters within the experimental limit of the particular experimental device. After completing the plasma diagnostic part using Langmuir probe, IA wave is excited in the plasma and measured its damping rate for various plasma conditions to verify the presence of hot electrons. It is observed that the waves suffer more damping with the decrease in the hot to cold electron density ratio, confirming the presence of hot electrons in the system.

\section*{Acknowledgements}
The authors would like to thank Mr G D Sarma for his technical support in carrying out the experiment. The authors also wish to acknowledge the valuable comments and ideas of Dr. Nipan Das, Dr. Partha Saikia and Ms. Jinti Barman at different phases of the experiments.

\end{document}